\documentclass{aa520}
\usepackage{graphicx}
\usepackage{txfonts}
\usepackage{color}
\usepackage{pstricks}

\begin{document}
\title{Representations of world coordinates in FITS}
\author{E. W. Greisen\inst{(1)} \and
        M. R. Calabretta\inst{(2)}}
\institute{National Radio Astronomy Observatory,
           P. O. Box O,
           Socorro, NM
           USA 87801-0387
\and       Australia Telescope National Facility,
           P. O. Box 76,
           Epping, NSW 1710 Australia}
%
% FITS keywords, no argument.
\newcommand{\keyw}[1]{\hbox{{\tt #1}}}
% FITS keywords, single italicized argument.
\newcommand{\keyi}[2]{\hbox{{\tt #1\hspace{1pt}}{$#2$}\/}}
\newcommand{\CRPIX}[1]{\keyi{CRPIX}{#1}}
\newcommand{\CDELT}[1]{\keyi{CDELT}{#1}}
\newcommand{\CRVAL}[1]{\keyi{CRVAL}{#1}}
\newcommand{\CTYPE}[1]{\keyi{CTYPE}{#1}}
\newcommand{\CUNIT}[1]{\keyi{CUNIT}{#1}}
\newcommand{\NAXIS}[1]{\keyi{NAXIS}{#1}}
\newcommand{\CROTA}[1]{\keyi{CROTA}{#1}}
% FITS keywords, double italicized arguments.
\newcommand{\keyii}[3]{\hbox{{\tt #1\hspace{1pt}{$#2$}\_{$#3$}}\/}}
\newcommand{\PC}[2]{\keyii{PC}{#1}{#2}}
\newcommand{\CD}[2]{\keyii{CD}{#1}{#2}}
\newcommand{\PV}[2]{\keyii{PV}{#1}{#2}}
\newcommand{\PS}[2]{\keyii{PS}{#1}{#2}}
% FITS keyvalue.
\newcommand{\keyv}[1]{\hbox{{\tt #1}}}
% older ones
\newcommand{\PCij}{\PC{i}{j}}
\newcommand{\CDij}{\CD{i}{j}}
\newcommand{\Ci}{{\it a\/}}
\newcommand{\NN}[1]{\noalign{\noindent{#1}}\noalign{\vspace{4pt}}}
\offprints{E. Greisen,\\\email{egreisen@nrao.edu}}
%
% Real date $Date: 2002/11/13 18:28:46 $
\date{Received 24 July 2002 / Accepted 9 September 2002}
\abstract{
The initial descriptions of the FITS format provided a simplified
method for describing the physical coordinate values of the image
pixels, but deliberately did not specify any of the detailed
conventions required to convey the complexities of actual image
coordinates.  Building on conventions in wide use within astronomy,
this paper proposes general extensions to the original methods for
describing the world coordinates of FITS data.  In subsequent
papers, we apply these general conventions to the methods by which
spherical coordinates may be projected onto a two-dimensional plane
and to frequency/wavelength/velocity coordinates.
}
\maketitle
\markboth{E. W. Greisen and M. R. Calabretta: Representations of world
          coordinates in FITS}{}

\section{Introduction}

\label{s:intro}

The Flexible Image Transport System, or FITS format, was first
described by Wells et al.~(\cite{kn:WGH}).  This
format is characterized by a fixed logical record length of 2880
bytes, and the use of an unlimited number of character-format
``header'' records with an 80-byte, keyword-equals-value substructure.
The header is followed by the header-specified number of binary data
records, which are optionally followed by extension records of the
specified length, but, at that time, of unspecified format.  Since
then, a number of authors have suggested various types of extensions
(e.g.~Greisen \&\ Harten \cite{kn:GH}; Grosb\o l et
al.~\cite{kn:GHGW}; Harten et al.~\cite{kn:HGGW}; Cotton et 
al.~\cite{kn:CTP}).  Because of its great flexibility, the FITS
format has been, and continues to be, very widely used in astronomy.
In fact, the FITS tape format was recommended (resolution C1) for use
by all observatories by Commission 5 at the 1982 meeting of the IAU at
Patras (\cite{kn:IAU}) and the General Assembly of the IAU adopted
(resolution R11) the recommendations of its commissions, including the
FITS resolution. A committee of the NASA/Science Office of Standards
and Technology has codified the current state of FITS into a document
which has been accepted as the official definition of the standard
(Hanisch et al.~\cite{kn:NOST}).

Wells et al.~(\cite{kn:WGH}) anticipated the need to specify the
physical, or world, coordinates to be attached to each pixel of an
{\it N\/}-dimensional image.  By {\em world coordinates}, we mean
coordinates that serve to locate a measurement in some 
multi-dimensional parameter space.  They include, for example, a
measurable quantity such as the frequency or wavelength associated
with each point in a spectrum or, more abstractly, the longitude and
latitude in a conventional spherical coordinate system which define a
direction in space.  World coordinates may also include enumerations
such as ``Stokes parameters'', which do not form an image axis in the
normal sense since interpolation along such axes is not meaningful.

Wells et al.~(\cite{kn:WGH}) viewed each axis of the image as having
a coordinate type and a reference point for which the pixel
coordinate, a coordinate value, and an increment were given.  Note
that this reference point was not required to occur at integer pixel
locations nor even to occur within the image.  An undefined
``rotation'' parameter was also provided for each axis. Since there
are, in general, more coordinates to be attached to a pixel than there
are ``real'' axes in the {\it N\/}-dimensional image, the convention
of declaring axes with a single pixel was also established in both
examples given by Wells et al.  The keywords defined were
\begin{center}\begin{tabular}{ll}
\CRVAL{n} & coordinate value at reference point \\
\CRPIX{n} & array location of the reference point in pixels\\
\CDELT{n} & coordinate increment at reference point \\
\CTYPE{n} & axis type (8 characters) \\
\CROTA{n} & rotation from stated coordinate type
\end{tabular}\end{center}
\noindent A list of suggested values for \CTYPE{n}\ was provided with
few of the details actually required to specify coordinates.  The
units were specified to be The International System of Units ``SI''
(meters, kilograms, seconds) with the addition of degrees for angles.

The simplicity of this initial description was deliberate.  It was
felt that a detailed specification of coordinate types was a lengthy
and complicated business, well beyond the scope intended for the
initial paper.  In addition, the authors felt that a detailed
specification would probably be somewhat controversial and thus likely
to compromise the possibility of wide-spread agreement on, and use of,
the basic structures of the format.  Hindsight also suggests that we
were rather naive at the time concerning coordinates and it is
fortunate that the detailed specification was postponed until greater
experience could be obtained.

The descriptions of coordinates in the initial FITS paper are simply
inadequate.  They provide no description of the meaning of the world
coordinates and suggest a rather incomplete list of coordinate types.
The use of a single rotation per axis cannot describe any general
rotation of more than two axes.

While participating in the development of the AIPS software package of
the National Radio Astronomy Observatory, Greisen (\cite{kn:G1},
\cite{kn:G2}) found it necessary to supply additional details to the
coordinate definitions for both spectral and celestial coordinates.
Since the latter have been widely used, a NASA-sponsored conference
held in January 1988 recommended that they form the basis for a more
general coordinate standard (Hanisch \&\ Wells~\cite{kn:HW}); such a
standard is described below.

The present work generalizes the set of world coordinate system (WCS)
FITS keywords with a view to describing non-linear coordinate 
systems and any parameters that they may require.  Alternate
keywords which should be supported are described.  It also addresses
the questions of units, multiple coordinate descriptions,
uncertainties in the coordinate values, and various other coordinate
related matters.  Conventions for attaching coordinate information
to tabular data are also described.  Paper II (Calabretta \&\ Greisen
\cite{kn:CG}) and Paper III (Greisen et al.~\cite{kn:GVCA})
extend these concepts to the ideal, but non-linear angular and
spectral coordinates used in astronomy.  Paper IV (Calabretta et
al.~\cite{kn:CVGTDAW}) then provides methods to describe the
distortions inherent in the image coordinate systems of real
astronomical data.  The complex questions related to time systems
and to other kinds of coordinates will be deferred.

\section{Basic concepts}

\subsection{Coordinate definition and computation}
\label{s:coordcomp}

\begin{figure}
   \centerline{
      \includegraphics[height=227pt]{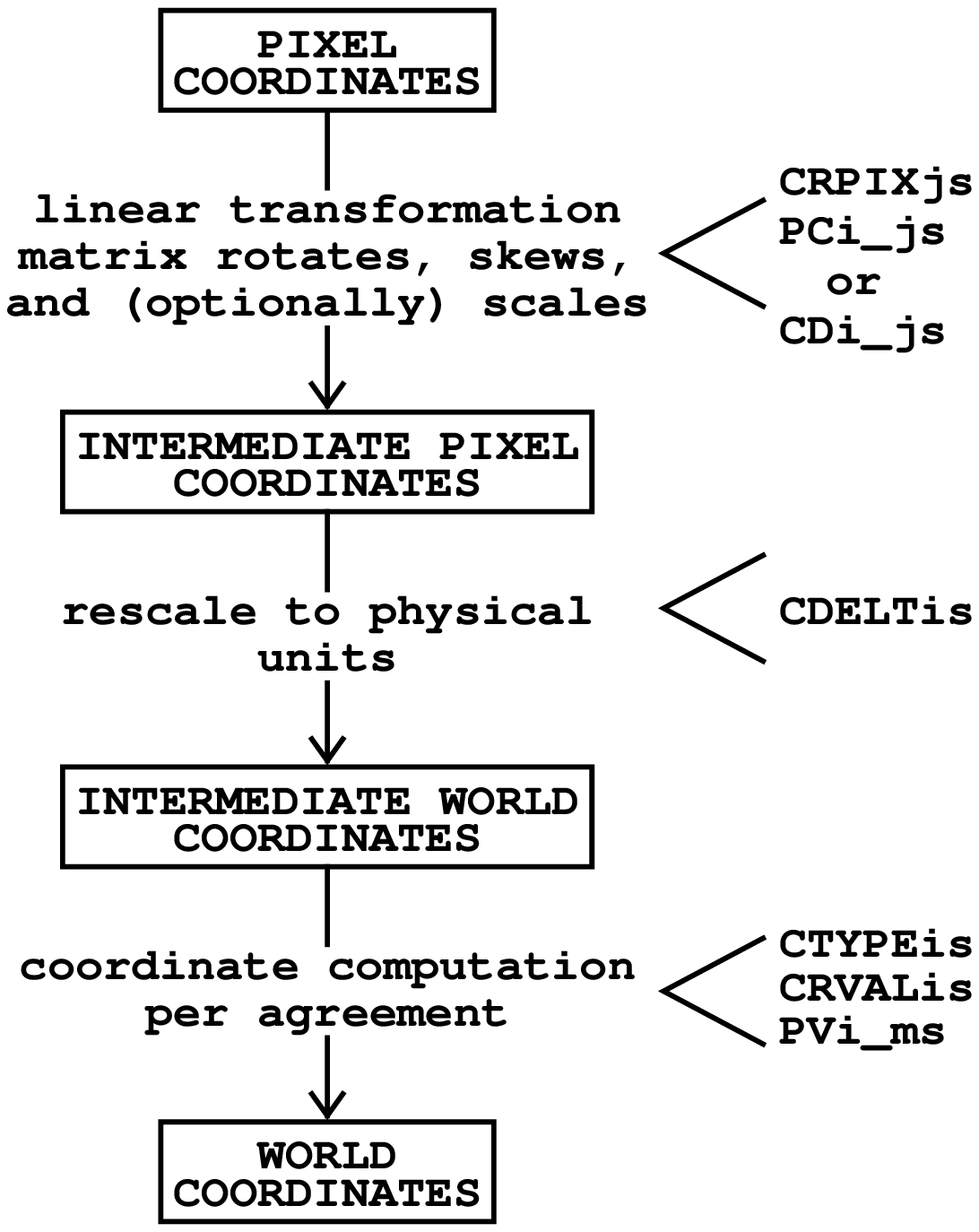}
      \begin{pspicture}(40pt,227pt)
        \psset{xunit=1pt,yunit=1pt,runit=1pt}
        \psset{linewidth=0.5}
        \psframe(0,0)(40,228)
        \psset{linewidth=0.8}
        \rput(20,214){$p_j$}
        \psframe(4,206)(36,225)
        \qline(20,206)(20,199)
        \rput(21,192){$r_j$}
        \rput(20,175){$m_{ij}$}
        \qline(20,170)(20,154)
        \psline(17,159)(20,154)(23,159)
        \rput(20,143){$q_i$}
        \psframe(4,134)(36,153)
        \qline(20,134)(20,121)
        \rput(20,116){$s_i$}
        \qline(20,111)(20,90)
        \psline(17,95)(20,90)(23,95)
        \rput(20,79){$x_i$}
        \psframe(4,70)(36,89)
      \end{pspicture}
  }
\caption[]{Conversion of pixel coordinates to world coordinates
shown as a multi-step process.  In the first step a linear
transformation is applied via matrix multiplication of the pixel
coordinate vector.  This linear transformation may be restricted to
the geometrical effects of rotation and skewness with scaling to
physical units deferred until the second step (\PCij\ plus \CDELT{i}\
formalism).  Alternatively, scaling may be applied via the matrix with
the second step omitted (\CDij\ formalism).  The final step applies a
possibly non-linear transformation to produce the final world
coordinates.  Although generic keywords for this step are defined in
this paper, the mathematical details, including the interpretation of
the {\sc intermediate world coordinates}, are deferred to later papers
which may also interpose additional steps in the algorithm chain.  For
later reference, the mathematical symbols associated with each step
are shown in the box at right.}
\label{fig:PROcess}
\end{figure}

In the current proposal, we regard the conversion of pixel coordinates
to world coordinates as a multi-step process.  This is illustrated
conceptually in Fig.~\ref{fig:PROcess}, which shows only the steps to
be discussed here. Later extensions may interpose additional steps as
required.  For example, Paper~II divides the final step into two with
the computation of intermediate spherical coordinates that are
subsequently converted to celestial coordinates via a spherical
rotation.  Paper~IV interposes optional distortion corrections between
the first and/or second steps of Fig.~\ref{fig:PROcess}.  Generally
these are intended to account for small residuals that cannot be
described by one of the standard world coordinate transformations.
These may arise in a variety of ways; naturally (e.g. aberration or
atmospheric refraction), via complex instrumental response functions
(e.g. data cubes produced by a Fabry-Perot interferometer for which
surfaces of constant wavelength are curved), by the intrinsic nature
of the system under study (e.g. surface coordinates of the asteroid
Eros), or as a result of instrumental peculiarities.

\subsubsection{Basic formalism}
\label{s:basics}

For all coordinate types, the first step is a linear transformation
applied via matrix multiplication of the vector of {\em pixel
coordinate} elements, $p_j$:

\noindent
\begin{equation}
   q_i = \sum_{j=1}^{N} m_{ij} (p_j - r_j) \, ,  \label{eq:pq}
\end{equation}

\noindent
where $r_j$ are the pixel coordinate elements of the reference point
given by the \CRPIX{j}.  Henceforth we will consistently
use $j$ for pixel axis indexing and $i$ for the world axes.

The $m_{ij}$ matrix is a non-singular square matrix of dimension $N
\times N$.  In the first instance, $N$ is given by the {\tt NAXIS}
keyword value, but this will be generalized with the introduction of
the {\tt WCSAXES} keyword in Sect.~\ref{s:dimens} so that the
dimension of the world coordinates need not match that of the data
array.

The elements, $q_i$, of the resulting {\em intermediate pixel
coordinate} vector are offsets, in dimensionless pixel units, from the
reference point along axes coincident with those of the {\em
intermediate world coordinates}. Thus the conversion of $q_i$ to the
corresponding intermediate world coordinate element, $x_i$, is a
simple scale:

\noindent
\begin{equation}
   x_i = s_i q_i \, .  \label{eq:qx}
\end{equation}

\noindent
We defer discussion of the encoding of $m_{ij}$ and $s_i$ as FITS
header cards to Sect.~\ref{s:matrixspec}.

The third step in the process of computing world coordinates depends
on the \CTYPE{i}.  For simple linear axes, the $x_i$ are interpreted
as offsets to be added to the coordinate value at the reference point
given by \CRVAL{i}.  Otherwise, the \CTYPE{i} define a function of the
$x_i$, the \CRVAL{i}, and, perhaps, other parameters that must be
established by convention and agreement.  Any \CTYPE{i}\ not covered
by convention and agreement shall be taken to be linear.

Non-linear coordinate systems will be signaled by \CTYPE{i}\ in ``4-3''
form: the first four characters specify the coordinate type, the fifth
character is a ``{\tt -}'', and the remaining three characters specify
an algorithm code for computing the world coordinate value, for
example {\tt 'ABCD-XYZ'}\@.  We explicitly allow the possibility that
the coordinate type may augment the algorithm code, for example {\tt
'FREQ-F2W'} and {\tt 'VRAD-F2W'} may denote somewhat different
algorithms (see Paper III)\@.   Coordinate types with names of less
than four characters are padded on the right with ``{\tt -}'', and
algorithm codes with less than three characters are padded on the
right with blanks, for example \hbox{{\tt 'RA---UV '}}\@.  However, we
encourage the use of three-letter algorithm codes.

Particular coordinate types and algorithm codes must be established by
convention.  Paper II constructs the framework for celestial
coordinate systems, and Paper III does so for spectral axes
(frequency-wavelength-velocity).  \CTYPE{i}\ values that are not in
``4-3'' form should be interpreted as linear axes.  It is possible that
there may be old FITS files with a linear axis for which \CTYPE{i}\
is, by chance, in 4-3 form.  However, it is very unlikely that it will
match a recognized algorithm code (use of three-letter codes will
reduce the chances).  In such a case the axis should be treated as
linear.

\subsubsection{Linear transformation matrix}
\label{s:matrixspec}

The proposal to replace the \CROTA{i}\ keywords of Wells et
al.~(\cite{kn:WGH}) with a general linear transformation matrix dates
from Hanisch \&\ Wells~(\cite{kn:HW}), although the details of its
implementation have undergone considerable evolution.  The main point
of divergence has been whether the matrix should completely replace or
simply augment the \CDELT{i}, but there are also important
differences relating to the default values of the matrix elements.

In defining a nomenclature which augments the \CDELT{i}\ we have
been guided by the following considerations: 

\begin{description}
\item[$\bullet$\ ] Where possible, standards should grow by
  generalizing existing usage rather than developing a separate 
  parallel usage.  Augmenting the existing \CDELT{i}\ with a
  separate transformation matrix that defaults to unity makes old
  headers equivalent to new ones that omit the keywords that define
  the transformation matrix.  In any case, the ``once FITS, 
  always FITS'' rule means that FITS readers must continue to
  interpret \CDELT{i}, so it makes sense for \CDELT{i}\ to
  retain its original function.   

\item[$\bullet$\ ] The transformation matrix then replaces the
  poorly defined \CROTA{i}\ with a nomenclature that allows
  for both skew and\eject fully general rotations.  We do not consider this
  replacement and the consequent deprecation of the \CROTA{i}\ to be
  inconsistent with the aim of generalizing existing usage 
  since, to our knowledge, the \CROTA{i}\ have had no formal
  definition other than the ``AIPS convention'' (Greisen~\cite{kn:G1},
  \cite{kn:G2}).  Both Wells et al.~(\cite{kn:WGH}) and Hanisch et
  al.~(\cite{kn:NOST}) state that ``users of this option should
  provide extensive explanatory comments.''  Paper~II describes
  the translation of the AIPS interpretation of \CROTA{i}\ to
  the new formalism. 

\item[$\bullet$\ ] A large fraction of WCS representations, perhaps
  the great majority, will not require the general linear 
  transformation.  FITS writers may continue to use \CDELT{i}, so
  FITS-writing software need not be rewritten to conform to the new
  formalism unless it needs the new features.

\item[$\bullet$\ ] The physical units of a general image may differ by
  many orders of magnitude, from frequencies of $10^{10}$ Hz (or more)
  to angles of $10^{-3}$ degrees (or less).  If the physical units
  enter into the linear transformation matrix, then the elements of
  that matrix will have very different magnitudes.  These issues pose
  difficulties both in computing and in understanding, and it may be
  simpler to defer application of physical units until the
  multiplication by \CDELT{i}.

\item[$\bullet$\ ] These difficulties are compounded when correcting
  for the distortions present in real instruments.  Paper IV will show
  that some instruments require distortion corrections before, and
  others after, the linear transformation matrix.  Such corrections
  may need to be expressed in terms directly related to pixel
  coordinates.  If the physical units enter into the linear
  transformation matrix, then the distortion corrections which come
  after the matrix would have to compensate for the physical units
  applied by it, effectively undoing and then redoing a multiplication
  by \CDELT{i}\@.  Furthermore, commensurability problems may
  arise when recording the maximum distortion correction for a WCS
  representation that mixes pre-, and post-corrections between axes. 

\item[$\bullet$\ ] A widely used formalism that discards \CDELT{i}\ 
  was developed by the Space Telescope Science Institute for the Hubble Space
  Telescope and was incorporated generally in the IRAF data analysis system.
  We therefore support this as an alternative method.
\end{description}

In the \PCij\ formalism, the matrix elements $m_{ij}$ are encoded in
\begin{center}
\begin{tabular}{l}
\noalign{\vspace{-5pt}}
\PCij\ \hspace{2em} (floating-valued)\\
\noalign{\vspace{-5pt}}
\end{tabular}
\end{center}
\noindent header cards, and $s_i$ as \CDELT{i}.  The $i$ and $j$
indices are used without leading zeroes, e.g. \keyw{PC1\_1} and
\keyw{CDELT1}.  The default values for \PCij\ are 1.0 for $i = j$ and
0.0 otherwise. The \PCij\ matrix must not be singular; it must have an
inverse. Furthermore, all \CDELT{i}\ must be non-zero.  In other
words, invertibility means that transformations which project from an
initial coordinate system of dimensionality {\tt WCSAXES} to a world
coordinate system of dimensionality less than {\tt WCSAXES} are
forbidden.

In the \CDij\ formalism Eqs.~(\ref{eq:pq}) and (\ref{eq:qx}) are
combined as
\noindent
\begin{equation}
   x_i = \sum_{j=1}^{N} (s_i m_{ij}) (p_j - r_j) \, ,  \label{eq:px}
\end{equation}
\noindent and the 
\begin{center}
\begin{tabular}{l}
\noalign{\vspace{-5pt}}
 \CDij\ \hspace{2em} (floating-valued)\\
\noalign{\vspace{-5pt}}
\end{tabular}
\end{center}
\noindent keywords encode the product $s_i m_{ij}$.  The $i$ and $j$
indices are used without leading zeroes, e.g. \keyw{CD1\_1}.  The
\CDij\ matrix must not be singular; it must have an inverse.
\CDELT{i}\ and \CROTA{i}\ are allowed to coexist with \CDij\ as an
aid to old FITS interpreters, but are to be ignored by new readers.
The default behavior for \CDij\ differs from that for \PCij; if one
or more \CDij\ cards are present then all unspecified \CDij\ default
to zero.  If no \CDij\ cards are present then the header is assumed to
be in \PCij\ form whether or not any \PCij\ cards are present since
this results in an interpretation of \CDELT{i}\ consistent with Wells
et al.~(\cite{kn:WGH}).

We specifically prohibit mixing of the \PCij\ and \CDij\ nomenclatures
in any FITS header data unit.  With this restriction, translation from
the \CDij\ formalism to the \PCij\ formalism is effected simply in the
keyword parsing stage of header interpretation;  the {\tt CD{\it
i}\_{\it j\/}} should be considered equivalent to the \PCij\, subject
to the considerations for default values noted above and with
\CDELT{i}\ set to unity.  Similarly, \CDij\ can be calculated from 
\PCij\  and \CDELT{i}\ following Eq.~\ref{eq:px}.

\subsubsection{Usage comments}

The proposal presented in this and the subsequent papers is not simple
and provides wide latitude for mistakes in describing the WCS and in
writing the FITS headers.  The result of an improperly described WCS
is simply undefined; it is the job of the FITS writer to produce a
correct description.  A simple error which could be made in a WCS
description, or with other parts of a header, is a repetition of
keywords with different values assigned to them.  If, for example,
{\tt BUNIT} were repeated with a new value, the data would have
unknown units but would be read correctly.  In binary tables, a second
value for \keyi{TFORM}{n} would cause the tabular data to be read
incorrectly.

This is a very general proposal!  The linear transformation matrix
allows for skew and fully general rotations.  The reader should note
that this allows dissimilar axes to be rotated into one another.  This
is meaningful in imaging; for example, one may wish to re-sample a
spectral-line cube from some special viewing angle in the three-space
of two celestial coordinates and one frequency coordinate.  Such
rotations are, however, forbidden into axes whose coordinate values
are, by convention, only integral.  Thus, if \CTYPE{i_0}\ indicates a
world coordinate of integral type, then row $i_0$ of the linear
transformation matrix must contain only one non-zero element, and this
would normally be 1.0 or at least integral.  Additionally, it must be
the only non-zero element in the column containing it.  The {\tt
STOKES} axis is one such coordinate; see Sect.~\ref{s:others}.

The linear transformation matrix could also be used to represent
images that have been transposed, e.g.~
\begin{displaymath}
  \mbox{{\tt PC}} = \left ( \begin{array}{ccc}
       0 & 1 & 0 \\ 0 & 0 & 1 \\ 1 & 0 & 0 \end{array} \right ) \, .
\end{displaymath}
This is a legal usage, but likely to confuse the reader.  In this
example, the FITS user will read in the header that the first element
of the world coordinate is \keyw{CTYPE1}, although this corresponds to
the second pixel axis.  Note that keywords \keyw{NAXIS1},
\keyw{CRPIX1}, \PC{i}{{\tt 1}}, and \CD{i}{{\tt 1}}, for example,
all refer to the first {\it pixel} axis in the image, while
\keyw{CTYPE1}, \keyw{CRVAL1}, \PC{{\tt 1}}{j}, \CD{{\tt 1}}{j},
and \keyw{CDELT1} all refer to the first world coordinate (``$q_1$''
and ``$x_1$'') element.  They must produce a correct result when
Eqs.~(\ref{eq:pq}) and (\ref{eq:qx}) or Eq.~(\ref{eq:px}) are
applied.  Thus, $x_1$ is of type \keyw{CTYPE1}\ even if it does not
change with {$p_1$} (to use the nomenclature of Eqs.~(\ref{eq:pq}),
(\ref{eq:qx}), and (\ref{eq:px})).  Therefore, it is good form to
transpose the header parameters along with the image so that the
on-diagonal terms in the transformation matrix predominate.  If the
{\tt PC} or {\tt CD} matrix is essentially diagonal, then the human
reader of the FITS header will have a better chance of understanding
the coordinate representation.

Equations (\ref{eq:pq}) and (\ref{eq:qx}) allow considerable
flexibility in the way the linear transformation is partitioned
between the \PCij\ and \CDELT{i}.  In the absence of any formal
constraints, the normal expectation would be that the \CDELT{i}\ be
used as scaling parameters as in the past.  This is straightforward if
\PCij\ is orthogonal, i.e.\ defines a pure rotation or simple
reflection, but not if it has an element of skewness.  In general, a
reasonable approach is to choose \CDELT{i}\ so that
\begin{equation}
   \sum_{j=1}^N \mbox{\PC{i}{j}}\,^2 = 1\, . \label{eq:constraint}
\end{equation}
for all $i$.  This normalization leaves orthogonal matrices unchanged,
and only slightly modifies matrices which are nearly orthogonal.  Note
that this is not the same as setting the determinant of the \PCij\
matrix to unity.  Note also that this constraint is optional and may
not be the most physically meaningful selection of the \PCij.  For
example, the conversion from the old \CROTA{i} nomenclature to the new
\PCij\ form described in Paper II does not satisfy this constraint
unless the \CDELT{i}\ are equal.

\subsubsection{Additional points}
\label{s:addpoints}

Note that integer pixel numbers refer to the center of the pixel
in each axis, so that, for example, the first pixel runs from pixel
number 0.5 to pixel number 1.5 on every axis.  Note also that the 
reference point location need not be integer nor need it even occur
within the image. The original FITS paper (Wells et al.~\cite{kn:WGH})
defined the pixel numbers to be counted from 1 to \NAXIS{j}\ 
($ \geq 1$) on each axis in a Fortran-like order as presented in the
FITS image.\footnote{This convention differs from the usual practice
in computer graphics where the pixels are counted from zero with pixel
centers as half integers (e.g.~Adobe Systems, Inc.~\cite{kn:ASI}). The
convention proposed here has been used extensively in FITS since the
format was invented and no argument has been advanced sufficiently
compelling to invalidate the many thousands of files written with that
convention.  Furthermore, we regard our image samples as ``voxels'' in
real physical space rather than pixels in two-dimensional display
space. As such, they may be viewed from any angle via transposition
and rotation.  The only point within the individual voxel that remains
invariant under those operations is its center and we argue,
therefore, that it is the center of the voxel which we should count.}
\eject

A WCS representation should be invertible in the sense that a
pixel coordinate, when transformed to a world coordinate, must be
uniquely recoverable from that world coordinate.  Note that this does
not require that each pixel coordinate in an image have a valid world
coordinate; as an example, pixel coordinates in the corner of a
Hammer-Aitoff projection of the full sky fall outside the map
boundary.  Nor need each valid world coordinate correspond to a pixel
coordinate; for example, the divergent poles of the Mercator
projection are inaccessible.  In practical terms, it means that two or
more different pixel coordinates should not map to the same world
coordinate, as exemplified by a cylindrical projection in which the
longitude spans more than $360\degr$.  Such coordinate systems, while
easy to construct, may be very difficult to interpret properly in all
respects, including that of drawing a coordinate grid.  Thus, while
they are not explicitly prohibited, it may be expected that general
WCS interpreting software may not handle them properly.

An additional convention is needed where non-linear axes must be
grouped, for example, the two axes which form a map plane.  In
general, all axes in the group must have identical algorithm codes and
a scheme must be established by convention for associating members of
the group and, if necessary, their order.  For example, Paper II
introduces the '{\it x}{\tt LON}/{\it x}{\tt LAT}' and '{\it yz}{\tt
LN}/{\it yz}{\tt LT}' conventions for associating longitude/latitude
coordinate pairs.  This should serve as a model for other cases.  Note
that grouping is not required for linear axes which are always
separable (in the mathematical sense) by means of a rotation or skew
applied via the linear transformation matrix.  

Some non-linear algorithms require parameter values, for example,
conic projections require the latitudes of the two standard parallels.
Where necessary, numeric parameter values will be specified via
\begin{center}
\begin{tabular}{l}
\noalign{\vspace{-5pt}}
\PV{i}{m} \hspace{2em} (floating-valued)\\
\noalign{\vspace{-5pt}}
\end{tabular}
\end{center}
\noindent keywords, where {\it i\/} is the intermediate world
coordinate axis number and {\it m\/} is the parameter number. Leading 
zeros are not allowed and {\it m\/} may have only those values defined
for the particular non-linear algorithm in the range 0 through 99
only.  There may also be auxiliary keywords which are required to
define, for example, the frames of reference used for celestial and
velocity coordinates.  

A few non-linear algorithms may also require character-valued
parameters, for example, table lookups require the names of the table
extension and the columns to be used.  Where necessary,
character-valued parameters will be specified via
\begin{center}
\begin{tabular}{l}
\noalign{\vspace{-5pt}}
\PS{i}{m} \hspace{2em} (character-valued)\\
\noalign{\vspace{-5pt}}
\end{tabular}
\end{center}
\noindent keywords, where {\it i\/} is the intermediate world
coordinate axis number and {\it m\/} is the parameter number. Leading
zeros are not allowed and {\it m\/} may have only those values defined
for the particular non-linear algorithm in the range 0 through 99
only.

The keywords proposed above and throughout the main body of this
manuscript apply to the relatively simple images stored in the main
FITS image data and in {\tt IMAGE} extensions Ponz et
al.~\cite{kn:PTM}).  When coordinates are used to describe image 
fragments in {\tt BINTABLE} extension tables (Cotton et
al.~\cite{kn:CTP}), additional nomenclature conventions are required.
These are described in Sect.~\ref{s:bint}.

\subsection{Coordinate dimensionality}
\label{s:dimens}

The number of world coordinate elements associated with a datum
can exceed the number of pixel coordinate elements which locate it in
the image data array.  For example, long-slit optical spectra are
naturally two-dimensional; normally the slit is aligned with one
(spatial) pixel axis and the dispersion coincides with the other
(spectral) pixel axis.  While the spectral representation is
straightforward -- one spectral pixel coordinate transforms to just
one spectral world coordinate (frequency, wavelength or velocity) --
the spatial representation would appear to be problematic.  Since the
slit can be oriented at any angle on the sky, the single pixel
coordinate which locates a datum along the length of the slit must
transform to two spatial (angular) coordinates, typically a right
ascension and a declination, neither of which need be constant.

In fact, this problem was solved very early in the history of FITS\@.
Wells et al.~(\cite{kn:WGH}) illustrate headers containing degenerate
axes, i.e.~axes having \NAXIS{j}~=~1, and this, combined with
the meaning assigned to \CROTA{i}\ by Greisen (\cite{kn:G1}),
provides a fully functional solution.  While not previously documented
outside the AIPS project, this solution is well known and has been
used extensively.  Basically the idea is simply to increment the
number of pixel coordinate elements as required by introducing
degenerate axes.

For the long slit example, we set {\tt NAXIS}~=~3 and \NAXIS{{\tt
3}}~=~1.  Supposing without loss of generality that \keyw{CTYPE1}\ is
the spectral axis, we represent \keyw{CTYPE2}\ as right ascension and
\keyw{CTYPE3}\ as declination.  \CROTA{i}\ in the original formulation
is here replaced by \PCij, so that pixel coordinates along the length
of the slit, $(p_2, p_3=1)$, transform to intermediate world
coordinates $(x,y)$.  Thus the slit's locus in the $xy$-plane is a
straight line whose orientation can be controlled via the \PCij\
matrix. Details of the transformation of $(x,y)$ to celestial
spherical coordinates are properly the subject of Paper~II\@. However,
given that the $xy$-plane is to be interpreted as the map plane of a 
spherical projection, it should be clear that rotating the slit in the
$xy$-plane via the \PCij\ matrix corresponds to rotating it on
the sky.  Paper~II discusses this long slit example in detail.

There is concern that requiring the use of degenerate pixel axes would
have severe repercussions for a significant fraction of existing
software programs which were not written to deal with such usage.  For
example, some software intended to handle two-dimensional images would
reject a FITS header with {\tt NAXIS}~=~3, even if the third axis is
degenerate.  At the same time, degenerate axes are a widespread and
natural representation for images in a multi-dimensional space.
Furthermore, as shown in Fig.~1 of Wells et al.~(\cite{kn:WGH}),
explicit specification of degenerate axes allows them to appear in any
order.  Such usage may facilitate image building and sub-imaging
operations.

To provide a solution for this world-coordinate dimensionality problem
that does not require the use of degenerate axes, we reserve the
keyword 
\begin{center}
\begin{tabular}{l}
\noalign{\vspace{-5pt}}
{\tt WCSAXES} \hspace{2em} (integer-valued)\\
\noalign{\vspace{-5pt}}
\end{tabular}
\end{center}
\noindent to specify the highest value of the index of any WCS keyword
in the header (i.e. \CRPIX{j}, \PCij\ or \CDij, \CDELT{i},\pagebreak\ \CTYPE{i},
\CRVAL{i}, or \CUNIT{i}).  The default value is the larger of {\tt
NAXIS} and the largest index of these keywords found in the FITS
header.  This keyword, if present, must precede all WCS keywords
(other than the \NAXIS{j}). The use of this keyword also solves the
problem posed by alternate axis descriptions (Sect.~\ref{s:alternate})
which may have an intrinsically different coordinate dimensionality.
In the slit example, an alternate description of the $x,y$ coordinates
on the detector has no use for a third axis.  It is a good idea to
provide a coordinate description, even if it is only a ``pixel'' axis,
for all array axes having more than one pixel.

There is debate within the community as to whether the official
definition of FITS (Hanisch et al.~\cite{kn:NOST}) prohibits the
occurrence of WCS-related keywords with indices greater than the value
of {\tt NAXIS}\@.  We make no claims one way or the other, but rather
assert that in order to accommodate WCS specifications whose
dimensionality exceeds {\tt NAXIS} without the use of degenerate
coordinate axes, such use must be allowed.  Consistent with Hanisch et
al.~(\cite{kn:NOST}), however, no \NAXIS{j}\ keywords may exist for $j
> $ {\tt NAXIS}\@.  Thus, calculations related to determining the
total length of the data array, which relies upon a product of the
\NAXIS{j}\ values, are unaffected.  Accordingly, all axes with axis
number greater than {\tt NAXIS} must be one pixel in length implicitly
rather than explicitly.

\subsection{Keyword value units}
\label{s:cunits}

The original FITS paper (Wells et al.~\cite{kn:WGH}) assumed that the
units along each axis could be implied simply by the contents of the
\CTYPE{i}\ keyword.  This has not turned out to be true in
general.  Therefore, we propose adding a new indexed, keyword
\begin{center}
\begin{tabular}{l}
\noalign{\vspace{-5pt}}
 \CUNIT{i} \hspace{2em} (character-valued)\\
\noalign{\vspace{-5pt}}
\end{tabular}
\end{center}
\noindent with which the units of \CRVAL{i}\ and \CDELT{i}\ may
be specified. Restrictions on the nature and range of units, if any,
will be determined by the agreements applying to the specific axes.
If they are not so limited, units should conform with the
recommendations of the IAU Style Manual (McNally~\cite{kn:IAUsm}).
Particular conventions for \CUNIT{i}\ values are discussed in
Sect.~\ref{s:unit}. Case will be significant in the values assigned to
\CUNIT{i}\ since, for example, it is necessary to represent both milli
and Mega prefixes for units such as {\tt mJy} and {\tt MJy}.  The
values assigned to \CUNIT{i}\ cannot exceed 68 characters, but may
well be longer than the 8 characters which has been a traditional --
but optional -- limit for character-valued but non-mandatory keywords.

\subsection{Keyword value defaults}
\label{s:defaults}

The original FITS paper also assumed that the coordinate keywords, if
present, would all be present and, therefore did not\pagebreak\  define defaults
for the standard keywords.  We therefore define the defaults to be
\begin{center}
\begin{tabular}{ll}
\noalign{\vspace{-5pt}}
{\tt WCSAXES} & {\tt NAXIS} or largest $i$ or $j$ \\
\CRVAL{i} & 0.0 \\
\CRPIX{j} & 0.0 \\
\CDELT{i} & 1.0 \\
\CTYPE{i} & {\tt ' '} (i.e.~a linear undefined axis) \\
\CUNIT{i} & {\tt ' '} (i.e.~undefined) \\
\PC{i}{j} & 1.0 when $i = j$ \\
\PC{i}{j} & 0.0 when $i \neq j$\\
\CD{i}{j} & 0.0 but see Sect.~\ref{s:matrixspec}\\
\noalign{\vspace{-5pt}}
\end{tabular}
\end{center}

\noindent These defaults provide the minimal amount of information
consistent with a real axis that is not fully described but will not
cause zero divides.  They assert that the pixel coordinate value
changes as the pixel number changes and that, by default, coordinate
values on the pixel axis depend upon that axis and only that axis.
The reference pixel is by default just off the data array which
satisfies the needs of some software systems and has the felicitous
result that the coordinate value of each pixel is its pixel number if
all of the keywords take their default values.  With these defaults, a
program may fill its coordinate arrays with usable, if uninteresting,
values before reading the FITS header, rather than constructing some
scheme that changes depending on the {\it absence} of keywords in the
FITS header.  Because default values were not defined from the
beginning and appear to be a source of confusion, we recommend that
FITS writers should endeavor always to write complete WCS
specifications and never to depend upon defaults.

\subsection{Alternate axis descriptions}
\label{s:alternate}

In some cases, an axis of an image may be described as having more
than one coordinate type.  An example of this would be the frequency,
velocity, and wavelength along a spectral axis (only one of which, of
course, could be linear).  One can also describe the position on a
CCD camera chip (or photographic plate) in meters as well as in
degrees on the sky.  To allow up to 26 additional descriptions of each
axis, we propose the addition of the following optional, but now
reserved, keywords defined in Table~\ref{ta:alternate},
where {\it j\/} and {\it i\/} are pixel and intermediate world
coordinate axis numbers, respectively, and \Ci\ is a character {\tt
A} through {\tt Z} specifying the coordinate version. The axis numbers
are restricted by this convention to the range 1--99 and the
coordinate parameter {\it m\/} is restricted to the range 0--99, both
with no leading zeros.  Note that the primary version of the
coordinate description is that specified with \Ci\ as the blank
character.  If an alternate coordinate description is specified, all
coordinate keywords for that version must be given even if they do not
differ from those of the primary version.  Rules for the default
values of alternate coordinate descriptions are the same as those for
the primary description.  The alternate coordinate descriptions are
computed in the same fashion as the primary coordinates.  The type of
coordinate depends on the value of \CTYPE{i\Ci}\ and may be linear in
one of the alternate descriptions and non-linear in another.
\begin{table}
\centering
\caption[]{Keywords with alternate axis descriptor codes.}
\vspace{5mm}
\begin{tabular}{ll}
\noalign{\vspace{-5pt}}
{\tt WCSAXES\Ci} & number of axes in WCS description \\
                      & (integer) \\
\CRVAL{i\Ci} & coordinate value at reference point \\
                      & (real floating) \\
\CRPIX{j\Ci} & pixel coordinate of the reference point \\
                      & (real floating) \\
\PC{i}{j\Ci}   & linear transformation matrix \\
                      & (real floating) \\
\CDELT{i\Ci} & coordinate increment \\
                      & (real floating) \\
\CD{i}{j\Ci} & linear transformation matrix
                        (with scale)\\ 
                      & (real floating) \\
\CTYPE{\Ci} & axis type \\
                      & (8 characters) \\
\CUNIT{i\Ci} & units of \CRVAL{i\Ci}\ and \CDELT{i\Ci} \\ 
                      & (character) \\
\PV{i}{m\Ci}         & coordinate parameter {\it m\/} \\
                      & (real floating) \\
\PS{i}{m\Ci}     & coordinate parameter {\it m\/} \\
                      & (character) \\
\noalign{\vspace{-5pt}}
\end{tabular}
\label{ta:alternate}
\end{table}

Alternate axis descriptions are optional, but may only be specified if
a primary axis description is specified.  The\pagebreak\ alternate version codes
are selected by the FITS writer; there is no requirement that the
codes be used in alphabetic sequence and no requirement that one
coordinate version differ in its parameter values from another.

An optional keyword
\begin{center}
\begin{tabular}{l}
\noalign{\vspace{-5pt}}
{\tt WCSNAME}{\Ci} \hspace{2em} (character-valued)\\
\noalign{\vspace{-5pt}}
\end{tabular}
\end{center}
\noindent is also defined to name, and otherwise document, the various
versions of world coordinate descriptions.  This keyword can be used
to give the user simple names by which to request the various versions
of the coordinates.  It may also be used, for example, to distinguish
coordinates used during data acquisition from those determined later
by astrometrically rigorous reductions.  It might also be used to
specify which are data pixels and which are calibration pixels in a
CCD image.

\subsection{Uncertainties in the coordinates}
\label{s:errors}

The coordinates of a pixel may not always be known exactly.  Instead,
they are often subject to both random, statistical errors and various
systematic errors.  The former are not particularly correlated between
pixels, whereas the latter may have a high degree of correlation
across the whole data set.  For example, single-dish radio images may
be accurate on a pixel-to-pixel basis to a fraction of an arcsec, but
have a 5--10 arcsec uncertainty in the reference point value.  Two
optional keywords are hereby reserved to specify these uncertainties
in coordinates.  They are
\begin{center}
\begin{tabular}{ll}
\noalign{\vspace{-5pt}}
{\tt CRDER{\it i\Ci}} & random error in coordinate \\
                      & (real floating) \\
{\tt CSYER{\it i\Ci}} & systematic error in coordinate \\
                      & (real floating) \\
\noalign{\vspace{-5pt}}
\end{tabular}
\end{center}
\noindent where both are given in units of \CUNIT{i\Ci} and
have default value zero.  They are understood to give a representative
average value of the error over the range of the coordinate in the
data file.  The total error in the coordinate would be given by
summing the two errors in quadrature.

The errors in actual coordinates may be very much more complex than
this simple representation.  In the most general case, one might
require, at each pixel, a covariance matrix to describe the dependence
of the uncertainty in one coordinate on the uncertainties in the
others.  Furthermore, the errors in one coordinate description may, or
may not, be completely predictable from those of an alternate
description.  Such usages, while perhaps important under some
circumstances, are well beyond the needs of most users and the scope
of this manuscript.

\section{Alternate FITS image representations: pixel list and vector
column elements\protect\footnotemark}\footnotetext{Contributed by
William Pence, Arnold Rots, and Lorella Angelini of the NASA Goddard
Space Flight Center, Greenbelt, MD 20771.}

\begin{table*}
\caption[]{Coordinate keywords for use in tables: the data type of the
table keyword matches that of the primary array keyword.  See
Sect.~\ref{s:keycons} for the definitions of the italicized
metasyntactic variables used below.}
\begin{center}
\setlength{\tabcolsep}{1.8em}
\protect\begin{tabular}{llllll}        
\hline\hline
Keyword\rule{0mm}{4mm} & Primary &\multicolumn{2}{c}{BINTABLE vector} &\multicolumn{2}{c}{Pixel List} \\
Description\rule[-2mm]{0mm}{3mm} & Array & primary & alternate & primary & alternate \\ 
\hline
Coordinate dimensionality\rule{0mm}{4mm}
                       & {\tt WCSAXES{\Ci}}    & \multicolumn{2}{c}{{\tt WCAX}{\it n\Ci}}
                                               & \multicolumn{2}{c}{--} \\
 Axis Type             & \CTYPE{i\Ci} & {\it i\/}{\tt CTYP}{\it n} & {\it i\/}{\tt CTY}{\it n\Ci}
                                               & {\tt TCTYP{\it n}}         & {\tt TCTY{\it n\Ci}} \\
 Axis Units            & \CUNIT{i\Ci} & {\it i\/}{\tt CUNI}{\it n} & {\it i\/}{\tt CUN}{\it n\Ci}
                                               & {\tt TCUNI{\it n}}         & {\tt TCUN}{\it n\Ci}\\
 Reference value       & \CRVAL{i\Ci} & {\it i\/}{\tt CRVL}{\it n} & {\it i\/}{\tt CRV}{\it n\Ci}
                                               & {\tt TCRVL{\it n}}         & {\tt TCRV{\it n\Ci}} \\
 Coordinate increment  & \CDELT{i\Ci} & {\it i\/}{\tt CDLT}{\it n} & {\it i\/}{\tt CDE}{\it n\Ci}
                                               & {\tt TCDLT{\it n}}         & {\tt TCDE{\it n\Ci}} \\
 Reference point       & \CRPIX{j\Ci} & {\it j\/}{\tt CRPX}{\it n} & {\it j\/}{\tt CRP}{\it n\Ci}
                                               & {\tt TCRPX{\it n}}         & {\tt TCRP{\it n\Ci}} \\
 Transformation matrix & \PC{i}{j\Ci} & \multicolumn{2}{c}{{\it ij\/}{\tt PC}{\it n\Ci}}
                                               & \multicolumn{2}{c}{{\tt TP{\it n}\_{\it k\Ci}}} \\
 Transformation matrix & {\tt CD{\it i}\_{\it j\Ci}} & \multicolumn{2}{c}{{\it ij\/}{\tt CD}{\it n\Ci}}
                                               & \multicolumn{2}{c}{{\tt TC{\it n}\_{\it k\Ci}}} \\
 Coordinate parameter  & \PV{i}{m\Ci}          & \multicolumn{2}{c}{{\it i\/}{\tt V{\it n}\_{\it m\Ci}}}
                                               & \multicolumn{2}{c}{{\tt TV{\it n}\_{\it m\Ci}}} \\
 Coordinate parameter array&  --               & \multicolumn{2}{c}{{\it i\/}{\tt V{\it n}\_X{\it\Ci}}}
                                               & \multicolumn{2}{c}{--} \\
 Coordinate parameter  & \PS{i}{m\Ci}          & \multicolumn{2}{c}{{\it i\/}{\tt S{\it n}\_{\it m\Ci}}}
                                               & \multicolumn{2}{c}{{\tt TS{\it n}\_{\it m\Ci}}} \\
 Coordinate name       & {\tt WCSNAME{\Ci}}    & \multicolumn{2}{c}{{\tt WCSN}{\it n\Ci}}
                                               & \multicolumn{2}{c}{{\tt TWCS}{\it n\Ci}} \\
 Random error          & {\tt CRDER{\it i\Ci}} & \multicolumn{2}{c}{{\it i\/}{\tt CRD}{\it n\Ci}}
                                               & \multicolumn{2}{c}{{\tt TCRD{\it n\Ci}}} \\
 Systematic error      & {\tt CSYER{\it i\Ci}} & \multicolumn{2}{c}{{\it i\/}{\tt CSY}{\it n\Ci}}
                                               & \multicolumn{2}{c}{{\tt TCSY{\it n\Ci}}} \\
 WCS cross-ref.~target & --                    & \multicolumn{2}{c}{{\tt WCST}{\it n\Ci}}
                                               & \multicolumn{2}{c}{--} \\
 WCS cross reference   & --                    & \multicolumn{2}{c}{{\tt WCSX}{\it n\Ci}}
                                               & \multicolumn{2}{c}{--} \\
 $\dag$ Coordinate rotation\rule[-2mm]{0mm}{3mm}
                       & \CROTA{i}    & {\it i\/}{\tt CROT}{\it n} &                             
                                               & {\tt TCROT{\it n}}         &                              \\
\hline
\end{tabular}
\end{center}
$\dag$ \CROTA{i}\ form is deprecated.  It may be used only when \PCij,
\PV{i}{m}, and \PS{i}{m}\ are not used and when {\it \Ci} is blank.
\label{ta:bintdef}
\end{table*}

\label{s:bint}

In addition to the image format discussed in the previous sections of
this paper (i.e.~an $N\/$-dimensional array in a FITS primary array or
FITS {\tt IMAGE} extension), there are two other FITS image
representations that are used commonly by the astronomical community
in binary tables extensions (Cotton et al.~\cite{kn:CTP}) in the
forms of:
\begin{enumerate}
\item
a multi-dimensional vector in a single element of a FITS binary
table,
\item
a tabulated list of pixel coordinates in a FITS ASCII or binary table,
and
\item
a combination of the two forms in a FITS binary table.
\end{enumerate}

The purpose of this section is to define a naming convention for the
coordinate system keywords to be used with these alternate image
formats.  Keywords specific to celestial coordinates will be treated
in Paper II and an example will be given.  Keywords specific to
spectral coordinates will be treated in a section of Paper III\@.
This general convention has been used for some time and is therefore
considered part of the full world coordinates convention.

The NOST (Hanisch et al.~\cite{kn:NOST}) standard provides that the
interpretation of raw field values found in any column $n$ of a FITS
table (either ASCII or binary) may be transformed into true physical
values by the presence of the keywords \keyi{TZERO}{n} and
\keyi{TSCAL}{n} for that column.  The tabular WCS keywords defined in
this section (and in the corresponding tabular keyword sections of
subsequent WCS papers) operate on the true physical values, not on the
raw field values.  Therefore any transformation specified by
\keyi{TZERO}{n} and \keyi{TSCAL}{n} is to be applied before these
tabular WCS computations.

\subsection{Multi-dimensional vector in a binary table}

A vector column in a binary table ({\tt BINTABLE}) extension can be
used to store a multi-dimensional image in each element (i.e.~each
row) of the column.  In the simple case in which all the images have
the same fixed size, the {\tt TDIM}{\it n\/} keyword can be used to
specify the dimensions.  In the more general case, a variable length
vector may be used to store different-sized images within the same
column.

Because two or more columns in a binary table can contain images, the
naming convention for these coordinate system keywords must encode the
column number containing the image to which the keyword applies as well
as the axis number within the image.  The naming convention described
here uses the keyword prefix to specify the axis number and the keyword
suffix to specify the column number containing the image (e.g.~the
{\tt 2CRVL15} keyword applies to the second axis of the image in
Col.~15 of the table).

\subsection{Tabulated list of pixels}

An image may also be represented as a list of $p_1, p_2, \ldots$ pixel
coordinates in a binary or ASCII table extension.  This representation
is frequently used in high-energy astrophysics as a way of recording
the position and other properties of individually detected photons.
This image format requires a minimum of $n$ table columns which give
the $p_1, p_2, \ldots, p_n$ (axes 1 through $n$) pixel coordinate of
the corresponding event in the virtual $n$-D image; any number of
other columns may be included in the table to store other parameters
associated with each event such as arrival time or photon energy.
This virtual image may be converted into a real image by computing the
$n$-dimensional histogram of the number of listed events that occur in
each pixel of the image (i.e.~the intensity value assigned to each
pixel ($p_1, p_2, \ldots, p_n$) of the image is equal to the number of
rows in the table which have axis 1 coordinate $= p_1$, axis 2
coordinate $= p_2$), etc.

A variation on this pixel list format may be used to specify explicitly
the intensity value of each image pixel.  This case requires at least
$n+1$ table columns which specify the axis 1 coordinate, the axis 2
coordinate, etc.~plus the value of the pixel at that coordinate.
In this representation each pixel coordinate would only be listed at
most once in the table; pixels with a value $= 0$ may be omitted
entirely from the table to conserve space.

Each axis of the image in this representation translates into a
separate column of the table, so the suffix of the coordinate system
keywords all refer to a column number rather than an axis number
(e.g.~the {\tt TCRP12} keyword applies to the coordinates listed in
the $12^{{\rm th}}$ column of the table).  This form of WCS keyword
may only be used with columns containing scalar values; the {\tt
BINTABLE} form must be used with columns containing more than one
value per table cell.

The presence of data from each column within a particular row of a
table implies an association of those data with each other.  However,
no formal method has previously been defined for identifying and
associating the columns which contain image pixel coordinates,
although informal conventions using new keywords have been used.  Past
practice has been to use distinctive column names
(e.g.~\keyi{TTYPE}{n} keywords with values of {\tt 'DETX'} and {\tt
'DETY'}, or {\tt 'X'} and {\tt 'Y'}) which a human interpreter may use
to form associations.  However, this is not generally suitable for
interpretation by software.
   
The keywords defined for pixel lists in Table~\ref{ta:bintdef}
partially remedy this by identifying the pixel coordinate columns.
For example, the presence of \keyi{TCTY}{n\Ci} in the header of a
binary table identifies column $n$ as containing a pixel coordinate
rather than, say, a pixel value.  In so doing it also identifies the
binary table as a pixel list.  Moreover, where a pixel list contains
multiple coordinate representations, the presence of a complete set of
{\tt TP{\it n}\_{\it k\Ci}} keywords would also provide a method of
associating the coordinate axes of each representation.

Thus, pending a formal solution of this problem, it is suggested that
a complete set of {\tt TP{\it n}\_{\it k\Ci}} (or {\tt TC{\it n}\_{\it
k\Ci}}) keywords be included in the pixel list header to define an
association of coordinate axes.  It should be noted that, while such
an association is unordered, this is not a concern for the computation
of world coordinates.

%Tables which are pixel lists often contain many
%columns of information which are associated with the other columns
%only by their presence within a particular row.  The WCS value
%associated with that column may be computed using only the value
%within that column and the keywords for that column.  We propose that
%the non-versioned (i.e.~\Ci\ = {\tt '\ '}) pixel list WCS keywords
%(e.g. \keyi{TCTYP}{n}) be used when no association between column
%values is required.  This allows, for example, the WCS structure to be
%used to calibrate numerous columns of instrument monitor data retained
%with the image data.   When two or more columns must be associated,
%since the computation of the WCS associated with those columns
%requires those two or more values, then versioned keywords
%(e.g.\keyi{TCTY}{n\Ci}) with the same, non-blank \Ci\ are to be used
%for those columns to signal the required association.  If there is an
%alternate WCS also requiring an association of more than one column
%for its computation, then that association would be signaled with a
%different, non-blank value for \Ci.  Any association described by the
%{\tt TP{\it n}\_{\it k\Ci}} (or {\tt TC{\it n}\_{\it k\Ci}}) keywords
%must not conflict with the associtions signaled by the use of
%non-blank values of \Ci.

\subsection{Keyword naming convention}
\label{s:keycons}

Table~\ref{ta:bintdef} lists the corresponding set of coordinate
system keywords for use with each type of FITS image representation.
The data type of the table keyword matches that of the corresponding
primary image keyword.  The allowed values for these keywords are
identical for all three types of images as defined in the main body of
this paper.  The old and now deprecated keyword \CROTA{i}\ has been
used with tables and is included since readers will need to understand
this keyword even if writers should no longer write it.  See Paper II
for a discussion of this point.  To support current usage, the
keywords are given in their current form to be used for the primary
coordinate representation ({\it \Ci} is blank) and a new form to
support the new capability to specify alternate coordinates for the
same axis ({\it \Ci} is {\tt A} through {\tt Z})\@.  For new keywords,
the two forms are identical and are shown in a single column midway
between the columns for old primary and new alternate WCS keywords.
The following notes apply to the naming conventions used in
Table~\ref{ta:bintdef}:

\begin{itemize}

\item
The $j, i$ prefix and suffix characters are integers referring to a
pixel and intermediate world coordinate axis number, respectively, of
the array.  When used as a keyword suffix the image dimension may
range from 1 to 99 with no leading 0, but when used as a prefix the
integer is limited to a single digit to conform to the 8-character
keyword name limit so the image may only contain up to 9 dimensions.

\item
\Ci\ is a 1-character coordinate version code and may be blank
(primary) or any single uppercase character from {\tt A} through {\tt
Z}\@.  

\item
$n$ and $k$ are integer table column number without any leading zeros
(1--999).

\item
$m$ is an integer between 0 and 99 with no leading zero giving
the coordinate parameter number.  $m$ cannot exceed 9 when $n$ exceeds
99, but see the following Section.

\item
The guidelines given Sect.~\ref{s:cunits} must be applied to the
the value of the \CUNIT{i\Ci} keyword and its derivatives.
In particular the value is restricted to {\tt 'deg'} when referring to
celestial coordinates; see Paper II\@.
\end{itemize}

\subsection{Multiple images and the ``Greenbank Convention''}
\label{s:GB}

In the case of the binary table vector representation, all the images
contained in a given column of the table may not necessarily have the
same coordinate transformation values.  For example, the pixel
location of the reference point may be different for each image/row in
the table, in which case a single {\tt 1CRP}{\it n\/} keyword in the
header is not sufficient to record  the individual value required for
each image.  In such cases, the keyword must be replaced by a column
with the same name (i.e.~{\tt TTYPE{\it m\/} = '1CRP{\it n\/}'}) which
can then be used to store the pixel location of the reference point
appropriate for each row of the table.  This convention for expanding
a keyword into a table column (or conversely, collapsing a column of
identical values into a single header keyword) is commonly known as
part of the ``Greenbank Convention'' for FITS keywords and is
illustrated in the example header shown in Paper II (Table 9)\@.

There are several restrictions which may be too limiting for the
parameters of certain types of coordinates and, in particular, for the
distortion parameters to be introduced in Paper IV\@.  The limitation
to 8 characters limits the number of columns to 999, the number of
axes to 9, and the number of parameters to as few as 10 (numbered
0 through 9, for column numbers exceeding 99).  To avoid this
difficulty, we introduce the concept of a parameter array as a single
column of a table.  All the parameters of a coordinate are given up to
the maximum dimension of the column (given by keyword {\tt TFORM}{\it
n\/}) with no omitted parameters.  Such parameter arrays are signaled
by replacing the {\tt \_{\it m\/}} in the table column name with {\tt
\_X}\@.

The Greenbank and parameter-array conventions are not needed with
pixel lists since they are used to represent a single image.

\subsection{Coordinate system cross-references}

While a coordinate representation may be shared amongst image arrays
within the same {\em column} of a binary table, it may also happen
that several image arrays within the same {\em row} of a binary table
must share the same coordinate representation.  For example, each row
of a table might store a raw optical spectrum, the corresponding sky
background spectrum, a flux-calibrated spectrum derived from these,
and a spectrum of the error in each channel.  It would not be
appropriate to coerce these into a 2-dimensional data array with a 
heterogeneous second axis, and in any case this would complicate the
addition or removal of spectra, say, as the result of data reduction.
It also may not be satisfactory simply to repeate the coordinate
description for each spectrum.  For example, it would be preferable to
apply a wavelength calibration to one shared representation rather
than several identical copies.

This situation is handled by introducing coordinate system
cross-references.  These apply only to binary tables containing
multiple image array columns, they are not relevant to primary image
arrays or to pixel lists which represent only a single data set.

Coordinate system cross-references allow an image array in one column
to reference the coordinate system defined for an image array in
another column.  The cross-reference is specified by the keyword pair
\begin{center}
\begin{tabular}{l}
\noalign{\vspace{-5pt}}
\keyi{WCST}{n\Ci} \hspace{2em} (character-valued)\\
\noalign{\vspace{-5pt}}
\end{tabular}
\end{center}
\noindent for the referred-to (target) coordinate system, and
\begin{center}
\begin{tabular}{l}
\noalign{\vspace{-5pt}}
\keyi{WCSX}{n\Ci} \hspace{2em} (character-valued)\\
\noalign{\vspace{-5pt}}
\end{tabular}
\end{center}
\noindent for the referring (cross-referencing) system, and these must
have identical, case-sensitive, keyword values.  \keyi{WCSX}{n\Ci}
must not be combined with any Table~\ref{ta:bintdef} coordinate
keywords that use the same alternate descriptor.  With regard to the
Greenbank convention (Sect.~\ref{s:GB}), when \keyi{WCST}{n\Ci} and/or
\keyi{WCSX}{n\Ci} are columns of the table, the scope of the
keyword(s) is limited to one row of the table.  Thus the same value of
\keyi{WCST}{n\Ci} and \keyi{WCSX}{n\Ci} may be reused in different rows.

On encountering \keyi{WCSX}{n\Ci}, FITS header-parsing software must
resolve the reference by searching for \keyi{WCST}{n\Ci} with the same
value, extracting the column number and alternate descriptor suffix
encoded in the keyword itself, and then searching for and loading the
associated coordinate keywords.

To continue the example above, suppose that Col.~12 contains a raw
optical spectrum for which the coordinate system is fully specified,
and that {\tt WCST12B} has been set to '{\tt XREF1}'.  Then a sky
background spectrum in Col.~13 might reference this coordinate
system by setting {\tt WCSX13A = 'XREF1'}.  In this case, Col.~13
must not contain any of the Table~\ref{ta:bintdef} keywords for
alternate descriptor {\tt \Ci\ = A}, although it might contain
keywords for some other value of \Ci.  The example illustrates that
the alternate descriptor suffixes for \keyi{WCST}{n\Ci} and
\keyi{WCSX}{n\Ci} need not match; the association is via the keyword
value alone.

\begin{table}
\centering
\caption[]{Characters \&\ strings allowed to denote mathematical
   operations.}
\renewcommand{\arraystretch}{1.30}
\begin{tabular}{ll}
\hline\hline
\multicolumn{1}{l}{String\rule[-2mm]{0mm}{6mm}} 
& \multicolumn{1}{l}{Meaning} \\ \hline
\verb+str1 str2+\rule{0mm}{4mm}& Multiplication\\
\verb+str1*str2+ & Multiplication\\
\verb+str1.str2+ & Multiplication\\
\verb+str1/str2+  & Division \\
\verb+str1**expr+&  Raised to the power \verb+expr+  \\
\verb+str1^expr+ &  Raised to the power \verb+expr+  \\
\verb+str1expr+  &  Raised to the power \verb+expr+ \\
\verb+log(str1)+   & Common Logarithm (to base 10) \\
\verb+ln(str1)+    & Natural Logarithm \\
\verb+exp(str1)+   & Exponential ($e^{\verb+str1+}$) \\
\verb+sqrt(str1)+\rule[-2mm]{0mm}{3mm}  & Square root \\
\hline
\end{tabular}
\label{ta:cmpunit}
\end{table}

\begin{table}
\centering
\caption[]{Prefixes for multiples \& submultiples.}
\renewcommand{\arraystretch}{1.30}
\begin{tabular}{llcllc}
\hline\hline
\multicolumn{1}{c}{Submult} 
& \multicolumn{1}{c}{Prefix} 
& \multicolumn{1}{c}{Char} 
& \multicolumn{1}{c}{Mult} 
& \multicolumn{1}{c}{Prefix\rule[-1mm]{0mm}{4mm}} 
& \multicolumn{1}{c}{Char} \\ \hline
\rule[0mm]{0mm}{4mm}%
$10^{-1}$  & deci  & \verb+d+ &$10$      & deca  & \verb+da+ \\
$10^{-2}$  & centi & \verb+c+ &$10^{2}$  & hecto & \verb+h+  \\
$10^{-3}$  & milli & \verb+m+ &$10^{3}$  & kilo  & \verb+k+  \\
$10^{-6}$  & micro & \verb+u+ &$10^{6}$  & mega  & \verb+M+  \\
$10^{-9}$  & nano  & \verb+n+ &$10^{9}$  & giga  & \verb+G+  \\
$10^{-12}$ & pico  & \verb+p+ &$10^{12}$ & tera  & \verb+T+  \\
$10^{-15}$ & femto & \verb+f+ &$10^{15}$ & peta  & \verb+P+  \\
$10^{-18}$ & atto  & \verb+a+ &$10^{18}$ & exa   & \verb+E+  \\
$10^{-21}$ & zepto & \verb+z+ &$10^{21}$ & zetta & \verb+Z+  \\
$10^{-24}$ & yocto & \verb+y+ &$10^{24}$ & yotta & \verb+Y+\rule[-2mm]{0mm}{2mm}\\
\hline
\end{tabular}
\label{ta:mulunit}
\end{table}

\begin{table}
\centering
\caption[]{IAU-recommended basic units.}
\renewcommand{\arraystretch}{1.30}
\begin{tabular}{llll}
\hline\hline
\multicolumn{1}{l}{Quantity} &
\multicolumn{1}{l}{Unit} & \multicolumn{1}{l}{Meaning
                        \rule[0mm]{0mm}{4mm}} 
         & \multicolumn{1}{l}{Notes} \\ 
& \multicolumn{2}{l}{String} &\rule[-2mm]{0mm}{4mm} \\ \hline
& & & \\
\multicolumn{4}{l}{\it SI base \& supplementary units} \\
length                & \verb+m+   & meter   & \\
mass                  & \verb+kg+  & kilogram & \verb+g+ gram okay\\
time                  & \verb+s+   & second  & \\
plane angle           & \verb+rad+ & radian  & \\
solid angle           & \verb+sr+  & steradian & \\
temperature           & \verb+K+   & kelvin  & \\
electric current      & \verb+A+   & ampere  & \\
amount of substance   & \verb+mol+ & mole    & \\
luminous intensity    & \verb+cd+  & candela & \\
& & & \\
\multicolumn{4}{l}{\it IAU-recognized derived units} \\
frequency             & \verb+Hz+  & hertz   & s$^{-1}$ \\
energy                & \verb+J+   & joule   & N m \\
power                 & \verb+W+   & watt    & J s$^{-1}$ \\
electric potential    & \verb+V+   & volt    & J C$^{-1}$ \\
force                 & \verb+N+   & newton  & kg m s$^{-2}$ \\
pressure, stress      & \verb+Pa+  & pascal  & N m$^{-2}$ \\
electric charge       & \verb+C+   & coulomb & A s \\
electric resistance   & \verb+Ohm+ & ohm     & V A$^{-1}$\\
electric conductance  & \verb+S+   & siemens & A V$^{-1}$ \\
electric capacitance  & \verb+F+   & farad   & C V$^{-1}$ \\
magnetic flux         & \verb+Wb+  & weber   & V s \\
magnetic flux density & \verb+T+   & tesla   & Wb m$^{-2}$ \\
inductance            & \verb+H+   & henry   & Wb A$^{-1}$ \\
luminous flux         & \verb+lm+  & lumen   & cd sr \\
illuminance           & \verb+lx+  & lux     & lm m$^{-2}$ \\
\hline
\end{tabular}
\label{ta:IAUunit}
\end{table}

\section{Specification of units}

\label{s:unit}

\begin{table*}
\caption[]{Additional allowed units.}
\begin{center}
\renewcommand{\arraystretch}{1.03}
\protect\begin{tabular}{lclll}
\hline\hline
\multicolumn{2}{l}{Quantity} &
\multicolumn{1}{l}{Unit} & \multicolumn{1}{l}{Meaning
                        \rule[0mm]{0mm}{4mm}} 
                        & \multicolumn{1}{l}{Notes} \\ 
& & \multicolumn{1}{l}{String}& &\rule[-2mm]{0mm}{4mm} \\ \hline
plane angle\rule{0mm}{4mm}
               &      & \verb+deg+      & degree of arc         
                      & $\pi/180$ rad \\
               &      & \verb+arcmin+   & minute of arc
                      & $1/60$ deg \\
               &      & \verb+arcsec+   & second of arc
                      & $1/3600$ deg \\
               &      & \verb+mas+     & milli-second of arc
                      & $1/3600000$ deg \\
time           &      & \verb+min+        & minute & \\
               &      & \verb+h+        & hour & \\
               &      & \verb+d+        & day 
                      & $86400$ s\\
               &$\dag$& \verb+a+        & year (Julian)
                      & $31557600$ s ~~(365.25 d), peta a
                       (\verb+Pa+) forbidden\\
               &$\dag$& \verb+yr+        & year (Julian)
                      & \verb+a+ is IAU-style \\
energy$^*$     &$\dag$& \verb+eV+        & electron volt
                      & $1.6021765\times10^{-19}$ J \\
               &$\ddag$& \verb+erg+        & erg
                      & $10^{-7}$ J \\
               &      & \verb+Ry+       & rydberg
                      & $\frac{1}{2}\left(\frac{2\pi e^2}{hc}\right)^2
                          m_ec^2  = 13.605692 $ eV \\
mass$^*$       &      &\verb+solMass+   & solar mass
                      & $1.9891\times10^{30}$ kg \\
               &      &\verb+u+         & unified atomic mass unit
                      &$1.6605387\times10^{-27}$ kg \\
luminosity     &      & \verb+solLum+   & Solar luminosity
                      & $3.8268\times10^{26}$ W \\
length         &$\ddag$& \verb+Angstrom+ & angstrom 
                      & $10^{-10}$ m \\
               &      & \verb+solRad+   & Solar radius
                      & $6.9599\times10^8$ m \\
               &      & \verb+AU+       & astronomical unit 
                      & $1.49598\times10^{11}$ m \\
               &      & \verb+lyr+      & light year
                      & $9.460730\times10^{15}$ m \\
               &$\dag$& \verb+pc+& parsec
                      & $3.0857\times10^{16}$ m \\
events         &      & \verb+count+ & count  & \\
               &      & \verb+ct+    & count  & \\
               &      & \verb+photon+ & photon  & \\
               &      & \verb+ph+     & photon  & \\
flux density   &$\dag$& \verb+Jy+   & jansky
                      & $10^{-26}$ W m$^{-2}$ Hz$^{-1}$ \\
               &$\dag$& \verb+mag+   & (stellar) magnitude &  \\
               &$\dag$& \verb+R+     & rayleigh & $10^{10}/(4\pi)$ photons\, m$^{-2}$\, s$^{-1}$\, sr$^{-1}$ \\
magnetic field &$\dag\ddag$& \verb+G+     & gauss
                      & $10^{-4}$ T \\
area           &      & \verb+pixel+ & (image/detector) pixel & \\
               &      & \verb+pix+   & (image/detector) pixel & \\
               &$\dag\ddag$& \verb+barn+  & barn
                      & $10^{-28}$ m$^{2}$ \\
\multicolumn{5}{c}{Miscellaneous ``units''\rule[-2mm]{0mm}{6mm}}\\
               &      & \verb+D+     & debye
                      & $\frac{1}{3}\times 10^{-29}$ C.m \\
               &      & \verb+Sun+   & relative to Sun
                      & e.g.~abundances \\
               &      & \verb+chan+  & (detector) channel & \\
               &      & \verb+bin+   & numerous applications 
                      &  (including the 1-d analogue of pixel) \\
               &      & \verb+voxel+ & 3-d analogue of pixel & \\
               &$\dag$& \verb+bit+   & binary information unit & \\
               &$\dag$& \verb+byte+  & (computer) byte & 8 bit \\
               &      & \verb+adu+   & Analog-to-digital converter & \\
               &      & \verb+beam+  & beam area of observation & 
                                       as in Jy/beam\\ 
\hline 
\end{tabular}
\end{center}
$\dag$ - addition of prefixes for decimal multiples and submultiples
     are allowed.\\ 
$\ddag$ - deprecated in IAU Style Manual (McNally~\cite{kn:IAUsm}) but
     still in use. \\
$*$ - conversion factors from CODATA Internationally recommended
     values of the fundamental physical constants 1998({\tt
     http://physics.\\nist.gov/cuu/Constants/})
\label{ta:extunit}
\end{table*}

Unless agreed otherwise, units should conform with the recommendations
of the IAU Style Manual (McNally~\cite{kn:IAUsm}).  Unfortunately,
this manual defines units as they would appear in a published document
rather than as they must appear in plain character form.  George \&\
Angelini (\cite{kn:GA}) and Ochsenbein et al.~(\cite{kn:OPK}) have
prepared detailed documents on this subject upon which the following
remarks are based.  In particular, the following tables are taken from
George \&\ Angelini's manuscript (with some changes and additions).
Readers should consult these references for examples and expanded
discussion. We allow the possibility that the interpretation of
\CUNIT{i\Ci}\ may depend on conventions established for the
\CTYPE{i\Ci}\ associated with them.  For example, units specified by
``s'' might mean SI seconds or seconds of sidereal time.

The basic units string, called {\tt str1} and {\tt str2} in
Table~\ref{ta:cmpunit}, is composed of a unit string taken from Col.~2
of the IAU-recognized units in Table~\ref{ta:IAUunit} or the extended
astronomical units in Table~\ref{ta:extunit}.  All units from the
former and selected units from the latter may be preceded, with no
intervening blanks by a single character (two for deca) taken from
Table~\ref{ta:mulunit} and representing scale factors mostly in steps
of $10^3$.   Compound prefixes (e.g., \verb+ZYeV+ for $10^{45}$ eV)
are prohibited.  A compound string may then be created from these
simple strings by one of the notations in  Table~\ref{ta:cmpunit}.   A
unit raised to a power is indicated by the unit string followed, with
no intervening blanks, by the optional symbols \verb+**+ or \verb+^+
followed by the power given as a numeric expression, called
\verb!expr! in Table~\ref{ta:cmpunit}.  The power may be a simple
integer, with or without sign, optionally surrounded by parentheses.
It may also be a decimal number (e.g., 1.5, .5) or a ratio of two
integers (e.g. 7/9), with or without sign, which are always surrounded
by parentheses.  Thus meters squared is indicated by \verb!m**(2)!,
\verb!m**+2!, \verb!m+2!, \verb!m2!, \verb!m^2!, \verb!m^(+2)!,
etc.~and per meter cubed is indicated by \verb!m**-3!, \verb!m-3!,
\verb!m^(-3)!, \verb!/m3!, and so forth. Meters to the three halves
may be indicated by \verb!m(1.5)!, \verb!m^(1.5)!, \verb!m**(1.5)!,
\verb!m(3/2)!, \verb!m**(3/2)!, and \verb!m^(3/2)!, but {\it not} by
\verb!m^3/2! or \verb!m1.5!.

Note that functions such as {\tt log} actually require dimensionless
arguments, so, by {\tt log(Hz)}, for example, we actually mean {\tt
log({\it x}/1Hz)}.  The final string to be given as the value of
\CUNIT{i\Ci}\ is the compound string, or a compound of compounds,
preceded by an optional numeric multiplier of the form {\tt
10**{\it k}}, {\tt 10\^{}{\it k}}, or {\tt 10}$\pm k$ where {\it k} is
an integer, optionally surrounded by parentheses with the sign
character required in the third form in the absence of parentheses.
FITS writers are encouraged to use the numeric multiplier only when
the available standard scale factors of Table~\ref{ta:mulunit} will
not suffice. Parentheses are used for symbol grouping and are strongly
recommended whenever the order of operations might be subject to
misinterpretation.  A blank character implies multiplication which can
also be conveyed explicitly with an asterisk or a period.  Therefore,
although blanks are allowed as symbol separators, their use is
discouraged.  Two examples are {\tt '10**(46)erg/s'} and {\tt
'sqrt(erg/pixel/s/GHz)'}.  Note that case is significant throughout.
The IAU style manual forbids the use of more than one solidus ({\tt
/}) character in a units string. In the present conventions, normal
mathematical precedence rules are assumed to apply, and we, therefore,
allow more than one solidus.  However, authors might wish to consider,
for example, {\tt 'sqrt(erg/(pixel.s.GHz))'} instead of the form given
previously.

\section{Additional matters}

\subsection{Image display conventions}

It is very helpful to adopt a convention for the display of images
transferred via the FITS format.  Many of the current image processing
systems have converged upon such a convention.  Therefore, we
recommend that FITS writers order the pixels so that the first pixel
in the FITS file (for each image plane) be the one that would be
displayed in the lower-left corner (with the first axis increasing to
the right and the second axis increasing upwards) by the imaging
system of the FITS writer.  This convention is clearly helpful in the
absence of a description of the world coordinates.  It does not
preclude a program from looking at the axis descriptions and
overriding this convention, or preclude the user from requesting a
different display.  This convention also does not excuse FITS writers
from providing complete and correct descriptions of the image
coordinates, allowing the user to determine the meaning of the image.
The ordering of the image for display is simply a convention of
convenience, whereas the coordinates of the pixels are part of the
physics of the observation.

\subsection{Units in comment fields}

If the units of the keyword value are specified in the comment of the
header keyword, it is {\it recommended} that the units string be
enclosed in square brackets at the beginning of the comment field,
separated from the slash (``{\tt /}'') comment field delimiter by a
single space character.  This widespread, but optional, convention
suggests that square brackets should be used in comment fields only
for this purpose.  Nonetheless, no software should depend on there
being units expressed in this fashion within a keyword comment, nor
should any software depend on any string within square brackets in a
comment field containing a proper units string.  This convention is
purely for the human reader, although software could be written which
would interpret the string only if present and of proper content. 
If there is an established convention for the units of a keyword, then
only those units may be used.  An example, using a non-standard
keyword, is\\
{\tt EXPTIME =                1200. / [s] exposure time in seconds}

\subsection{Tables}

Binary extension tables (Cotton et al.~\cite{kn:CTP}) use the
\keyw{NAXIS2} keyword, invented for simple images, to specify the
number of rows in a table.  It has been suggested that, if the rows of
the table are regularly spaced in some world coordinate, that world
coordinate could be described with the other axis-2 keywords such as
\keyw{CTYPE2\Ci}, \keyw{CRVAL2\Ci}, \keyw{CDELT2\Ci}, and
\keyw{PC2\_2\Ci}.  Since we know of no software system using this
convention with these general keywords, we deprecate the suggestion.
There are very powerful and useful general operators such as sorting,
editing, and concatenation which alter the value of the row number and
therefore corrupt the value of the implied coordinate.  If FITS were
solely used as an interchange mechanism, these operators would not be
relevant. But FITS is now used as the internal format of several
software systems for which the general operators are important.
Initial row number can be recorded as a column in tables and
associated with a physical coordinate via keywords described in
Sect.~\ref{s:bint}.

\subsection{Conventional coordinate types}

\begin{table}
\centering
\caption[]{Conventional Stokes values.}
\renewcommand{\arraystretch}{1.30}
\begin{tabular}{rcl}
\hline\hline
Value & Symbol & Polarization \\
\hline
   1  & I      & Standard Stokes unpolarized \\
   2  & Q      & Standard Stokes linear \\
   3  & U      & Standard Stokes linear \\
   4  & V      & Standard Stokes circular \\
  -1  & RR     & Right-right circular \\
  -2  & LL     & Left-left circular \\
  -3  & RL     & Right-left cross-circular \\
  -4  & LR     & Left-right cross-circular \\
  -5  & XX     & X parallel linear\\
  -6  & YY     & Y parallel linear\\
  -7  & XY     & XY cross linear\\
  -8  & YX     & YX cross linear\\
\hline
\end{tabular}
\label{ta:Stokes}
\end{table}

\label{s:others}

In the first FITS paper, Wells et al.~(\cite{kn:WGH}) listed a number
of ``suggested values'' for \CTYPE{i}.  Two of these have the
attribute that they can assume only integer coordinate values and that
the meaning of these integers is only by convention.  These two axis
types are in wide-spread use and we wish to repeat their definition
here to extend their definitions and to reserve their names and
meanings.

The first conventional coordinate is \CTYPE{i\Ci}\ {\tt = 'COMPLEX'}
to specify complex valued data.  FITS data are limited to a single 
real number value at each pixel making this axis necessary to
represent data which are weighted complex numbers.  Conventional
values of 1 for the real part, 2 for the imaginary part, and 3 for a
weight (if any) have been widely used.

The second conventional coordinate is \CTYPE{i\Ci}\ {\tt = 'STOKES'}
to specify the polarization of the data.  Conventional values, their
symbols, and polarizations are given in Table~\ref{ta:Stokes}.

\begin{table*}
\caption[]{Coordinate keywords: see also Table~\ref{ta:bintdef}
 for alternate types used in binary tables.}
\renewcommand{\arraystretch}{1.30}
\begin{center}
\protect\begin{tabular}{llllll}
\hline\hline
 Keyword & Type & Sect. & Use & Status & Comments \\ 
\hline
{\tt WCSAXES\Ci}      & integer  & \ref{s:dimens} & WCS dimensionality & new
                      & allows WCS specification for degenerate axes, \\
                  & & & & & to be explicit rather than implicit.\\
\CRVAL{i\Ci}  & floating &\ref{s:basics}  & value at reference point & extended
                      & meaning of reference point forced by coord.~type. \\ 
\CRPIX{j\Ci}  & floating &\ref{s:basics}  & pixel of reference point & extended
                      & meaning of reference point forced by coord.~type. \\ 
\CDELT{i\Ci}  & floating &\ref{s:basics}  & increment at ref.~point & retained
                      & meaning of increment clarified. \\
\CROTA{i}     & floating &\ref{s:intro}   & rotation at ref.~point & deprecated
                      & replaced by \PCij\ and \CDij. \\
\CTYPE{i\Ci}  & character &\ref{s:basics}  & coord./algorithm type& extended
                      & Non-linear types have ``4-3'' form: characters 1--4\\
                  & & & & & specify the coordinate type, character 5 is ``{\tt -}'',\\
                  & & & & & and characters 6--8 specify an algorithm code\\
                  & & & & & for computing the world coordinate value; \\
                  & & & & & case dependent.\\
\CUNIT{i\Ci} & character & \ref{s:cunits} & units of coord.~values & new
                      & case dependent, allowed values and combinations \\
                  & & & & & are described in Sect.~\ref{s:unit}. \\
\PC{i}{j\Ci}          & floating & \ref{s:matrixspec} & transformation matrix & new
                      & linear conversion of pixel number to pixels along \\
                  & & & & & coordinate axes; default $= 0\;\; (i \neq j),\;\;\; = 1\;\; (i = j)$.\\ 
{\tt CD{\it i}\_{\it j\Ci}} & floating & \ref{s:matrixspec} & transformation matrix & new
                      & linear conversion of pixel number to relative \\
                  & & & & & coordinates; default all 0 if any given,\\
                  & & & & & else \PCij\ applies.\\ 
\PV{i}{m\Ci}         & floating & \ref{s:addpoints} & parameter {\it m\/} & new
                      & parameters required in some coordinate types;  \\
                  & & & & & defaults are algorithm-specific; see Papers II and\\
                  & & & & &  III for usage examples and default conventions\\
\PS{i}{m\Ci}    & character & \ref{s:addpoints} & parameter {\it m\/} & new
                      & parameters required in some coordinate types;  \\
                  & & & & & defaults are algorithm-specific; see Paper III\\
                  & & & & & for usage example and default conventions\\
{\tt WCSNAME{\it \Ci}} & character & \ref{s:alternate} & coord.~version name & new
                      & optional documentation \\
{\tt CRDER{\it i\Ci}} & floating & \ref{s:errors} & random error & new
                      & uncertainty in coordinate due to random errors;\\
                   & & & & & default $= 0$ \\
{\tt CSYER{\it i\Ci}} & floating & \ref{s:errors} & systematic error & new
                      & uncertainty in coordinate due to systematic\\
                   & & & & & errors; default $= 0$ \\
\hline
\end{tabular}
\end{center}
where {\it j\/} is a pixel axis number 1 through 99, {\it i\/} is an
intermediate world coordinate axis number 1 through 99, {\it m\/} is a
parameter number 0 \linebreak{2} through 99, and \Ci\ is a coordinate
description version character blank and {\tt A} through {\tt Z}\@.
\PC{i}{j\Ci} and \CD{i}{j\Ci} may not occur in the same header data
unit\@.
\label{ta:keyword}
\end{table*}

\section{Header construction example}
\label{sec:construct}

A simple header construction example based on Einstein's Special Theory of
Relativity (\cite{kn:E}) will serve to illustrate the formalism
introduced in this paper.  We will construct dual coordinate
representations, the first for the rest frame and the second for an
observer in uniform motion.

Suppose we have a data cube that, in the rest frame, has the following simple
header containing two spatial axes and one temporal axis:

\noindent
\begin{displaymath}
   \begin{array}{rclrcl}
      \keyw{NAXIS}   & = & \keyv{3} \, , &
      \keyw{CTYPE1}  & = & \keyv{'X'} \, , \\
      \keyw{NAXIS1}  & = & \keyv{2048} \, , &
      \keyw{CTYPE2}  & = & \keyv{'Y'} \, , \\
      \keyw{NAXIS2}  & = & \keyv{2048} \, , &
      \keyw{CTYPE3}  & = & \keyv{'TIME'} \, , \\
      \keyw{NAXIS3}  & = & \keyv{128} \, , &
      \keyw{CRVAL1}  & = & \keyv{0.0} \, , \\
      \keyw{CRPIX1}  & = & \keyv{1024.5} \, , &
      \keyw{CRVAL2}  & = & \keyv{0.0} \, , \\
      \keyw{CRPIX2}  & = & \keyv{1024.5} \, , &
      \keyw{CRVAL3}  & = & \keyv{0.0} \, , \\
      \keyw{CRPIX3}  & = & \keyv{64.5} \, , &
      \keyw{CUNIT1}  & = & \keyv{'km'} \, , \\
      \keyw{CDELT1}  & = & \keyv{3.0} \, , &
      \keyw{CUNIT2}  & = & \keyv{'km'} \, , \\
      \keyw{CDELT2}  & = & \keyv{3.0} \, , &
      \keyw{CUNIT3}  & = & \keyv{'us'} \, , \\
      \keyw{CDELT3}  & = & \keyv{10.0} \, , &
      \keyw{WCSNAME} & = & \keyv{'Rest frame'} \, .
   \end{array}
\end{displaymath}

\noindent
This describes three linear coordinate axes with the reference point in the
middle of the data cube.

The spatial and temporal coordinates measured by an observer moving with
uniform velocity $v$ in the $+x$ direction are related to the rest coordinates
by the Lorentz transformation:

\noindent
\begin{eqnarray*}
   x' & = & \gamma (x - vt)     \, , \\
   y' & = & y                   \, , \\
   t' & = & \gamma (t - vx/c^2) \, , \\
\noalign{\noindent{where}}
\noalign{\vskip 4pt}
   \gamma & = & 1 / \sqrt{1 - v^2/c^2} \, ,
\end{eqnarray*}

\noindent
and $c$ is the velocity of light.  Time in each system is measured from the
instant when the origins coincide.  From the above header we have

\noindent
\begin{eqnarray*}
   x & = & s_1 (p_1 - r_1) \, , \\
   y & = & s_2 (p_2 - r_2) \, , \\
   t & = & s_3 (p_3 - r_3) \, ,
\end{eqnarray*}

\noindent
where, $r_j$ and $s_i$ are given by \CRPIX{j} and \CDELT{i}.  Thus

\noindent
\begin{eqnarray*}
   x' & = & \gamma s_1 (p_1 - r_1) - \gamma v s_3 (p_3 - r_3) \, , \\
   y' & = & s_2 (p_2 - r_2) \, , \\
   t' & = & \gamma s_3 (p_3 - r_3) - \gamma v/c^2 s_1 (p_1 - r_1) \, .
\end{eqnarray*}

\noindent
This set of equations may be rewritten to make the scales, \CDELT{i}, the same
as the rest frame header:

\noindent
\begin{eqnarray*}
   x' & = & s_1 [ \gamma (p_1 - r_1) - \gamma v s_3/s_1 (p_3 - r_3) ] \, , \\
   y' & = & s_2 [ (p_2 - r_2) ] \, , \\
   t' & = & s_3 [ \gamma (p_3 - r_3) - \gamma v/c^2 s_1/s_3 (p_1 - r_1) ] \, .
\end{eqnarray*}

\noindent
Using character ``\keyv{V}'' as the alternate representation
descriptor, \Ci, for the relatively moving frame, we have

\noindent
\begin{displaymath}
   \begin{array}{rclrcl}
      \keyw{CRPIX1V}  & = & \keyv{1024.5} \, , &
      \keyw{CTYPE1V}  & = & \keyv{'X'} \, , \\
      \keyw{CRPIX2V}  & = & \keyv{1024.5} \, , &
      \keyw{CTYPE2V}  & = & \keyv{'Y'} \, , \\
      \keyw{CRPIX3V}  & = & \keyv{64.5} \, , &
      \keyw{CTYPE3V}  & = & \keyv{'TIME'} \, , \\
      \keyw{PC1\_1V}  & = & \hphantom{-}\gamma \, , &
      \keyw{CRVAL1V}  & = & \keyv{0.0} \, , \\
      \keyw{PC1\_3V}  & = &     - \gamma v/\sigma \, , &
      \keyw{CRVAL2V}  & = & \keyv{0.0} \, , \\
      \keyw{PC3\_1V}  & = &     - \gamma \sigma v/c^2 \, , &
      \keyw{CRVAL3V}  & = & \keyv{0.0} \, , \\
      \keyw{PC3\_3V}  & = & \hphantom{-}\gamma \, , &
      \keyw{CUNIT1V}  & = & \keyv{'km'} \, , \\
      \keyw{CDELT1V}  & = & \keyv{3.0} \, , &
      \keyw{CUNIT2V}  & = & \keyv{'km'} \, , \\
      \keyw{CDELT2V}  & = & \keyv{3.0} \, , &
      \keyw{CUNIT3V}  & = & \keyv{'us'} \, , \\
      \keyw{CDELT3V}  & = & \keyv{10.0} \, , &
      \keyw{WCSNAMEV} & = & \keyv{'Moving frame'} \, .
   \end{array}
\end{displaymath}

\noindent
Note that the elements of the \PC{i}{j} matrix are all dimensionless,
$\sigma = s_1/s_3 = 3 \times 10^{8} {\rm m}\,{\rm s}^{-1}$ having the
dimensions of a velocity.  However, in this instance we have seen fit
not to apply the strictures of Eq.~(\ref{eq:constraint}) in
normalizing the matrix.  In fact, in Minkowski space-time the concept
of ``distance'', on which Eq.~(\ref{eq:constraint}) relies, differs
from the Euclidean norm, the invariant being
\noindent
\begin{eqnarray}
   d & = & \sqrt{x^2 + y^2 + z^2 - c^2t^2} \, , \nonumber
\end{eqnarray}
\noindent
so one may query the fundamental validity of Eq.~(\ref{eq:constraint})
in this case.  However, the intent of that equation is well served
since \PC{i}{j} and \CDELT{i}\ are divided in a physically meaningful
way, especially considering that $\gamma$ is often close to unity so
that \PC{i}{j} is approximately the unit matrix. Nevertheless, the
appearance of the factor $\gamma$ in each of the elements of \PC{i}{j}
suggests the factorization 
\noindent
\begin{displaymath}
   \begin{array}{rclrcl}
      \keyw{PC1\_1V}  & = & \keyv{1} \, , &
      \keyw{CDELT1V}  & = &     3.0 \gamma \, , \\
      \keyw{PC1\_3V}  & = &     - v / \sigma \, , &
      \keyw{CDELT2V}  & = & \keyv{3.0} \, , \\
      \keyw{PC3\_1V}  & = &     - v/c^2 \sigma \, , &
      \keyw{CDELT3V}  & = &     10.0 \gamma \, , \\
      \keyw{PC3\_3V}  & = & \keyv{1} \, ,
   \end{array}
\end{displaymath}
\noindent
and indeed this does also have a physically meaningful interpretation in that
the scales are dilated by the Lorentz-Fitzgerald contraction factor, $\gamma$.

\section{Summary}

\label{s:summary}

The changes to FITS-header keywords are summarized in
Table~\ref{ta:keyword}.  As described in Paper II, for one purpose,
\CROTA{i}\ may in some cases be used instead of the new
keywords so that the coordinate information may be understood by
software systems which have yet to be converted to these new
conventions.

\begin{acknowledgements}
The authors would particularly like to thank Steve~Allen (University of
California, Lick Observatory) and Patrick~Wallace (U.K. Starlink) who
provided valuable encouragement, feedback and suggestions over the
long period of this work's development.  We are grateful for the
assistance provided by Bob Hanisch (Space Telescope Science Institute)
in bringing the manuscript to its final form. We thank William Pence,
Arnold Rots, and Lorella Angelini (NASA Goddard Space Flight Center)
for contributing the original text of Sect.~\ref{s:bint} and for
numerous other suggestions and comments.  We also thank Ian George and
Lorella Angelini (NASA Goddard Space Flight Center) for contributing 
the text that formed the basis of Sect.~\ref{s:unit} and Francois
Ochsenbein (Observatoire Astronomique, Strasbourg) for valuable
comments on that and other sections.

The authors also thank the following for comments and suggestions:
Lindsey~Davis, Brian~Glendenning, Doug~Mink, Jonathan~McDowell,
Tim~Pearson, Barry~Schlesinger, William~Thompson, Doug~Tody,
Francisco~Valdes, and Don~Wells.

The National Radio Astronomy Observatory is a facility of the (U.S.)
National Science Foundation operated under cooperative agreement by
Associated Universities, Inc.  

The Australia Telescope is funded by the Commonwealth of Australia for
operation as a National Facility managed by CSIRO\@.
\end{acknowledgements}

\end{document}